\documentclass[twocolumn,prb,superscriptaddress,aps,10pt,bibnotes,amsfonts,amsmath,amssymb,floatfix]{revtex4-2} 

\usepackage{graphicx}
\usepackage{ifthen}
\usepackage{xcolor}
\usepackage{bm}
\usepackage{braket}
\usepackage{multirow}
\usepackage{hyperref}
\usepackage{soul}
\usepackage{svg}
\usepackage{ulem}

\hypersetup{
 pdfnewwindow=true, colorlinks=true,
 linkcolor=blue, anchorcolor=blue,
 citecolor=blue, filecolor=blue,
 menucolor=blue, urlcolor=blue}

\definecolor{darkgreen}{RGB}{0, 136, 58}

\def\bk{\mathbf{k}}

\def\bq{\mathbf{q}}
\def\o{\omega}
\def\d{\delta}
\def\g{\gamma}
\def\ve{\varepsilon}
\def\ef{\ve_{\rm F}}
\def\dd{\mathrm{d}}

\begin{document}

\title{Phonon-limited carrier transport in the Weyl semimetal TaAs}

\author{Zhe Liu}
\affiliation{Department of Physics, Applied Physics and Astronomy, Binghamton University-SUNY, Binghamton, New York 13902, USA}
\altaffiliation{Equal Contribution}
\author{Shashi B. Mishra}
\affiliation{Department of Physics, Applied Physics and Astronomy, Binghamton University-SUNY, Binghamton, New York 13902, USA}
\altaffiliation{Equal Contribution}
\author{Jae-Mo Lihm}
\affiliation{European Theoretical Spectroscopy Facility and Institute of Condensed Matter and Nanosciences (IMCN), Universit\'e catholique de Louvain (UCLouvain), Belgium}
\affiliation{WEL Research Institute, avenue Pasteur 6, 1300 Wavre, Belgium}
\author{Samuel Ponc\'{e}}
\affiliation{European Theoretical Spectroscopy Facility and Institute of Condensed Matter and Nanosciences (IMCN), Universit\'e catholique de Louvain (UCLouvain), Belgium}
\affiliation{WEL Research Institute, avenue Pasteur 6, 1300 Wavre, Belgium}
\author{Elena R. Margine}
\email{rmargine@binghamton.edu}
\affiliation{Department of Physics, Applied Physics and Astronomy, Binghamton University-SUNY, Binghamton, New York 13902, USA}


\begin{abstract}
Topological Weyl semimetals represent a novel class of quantum materials that exhibit remarkable properties arising from their unique electronic structure. 
In this work, we employ state-of-the-art \textit{ab initio} methods to investigate the role of the electron-phonon interactions on the charge transport properties of TaAs. 
Our calculations of the temperature-dependent electrical conductivity with the iterative Boltzmann transport equation show excellent agreement with experimental measurements above 100~K. 
Extending the analysis to doped systems, we demonstrate that even small shifts in the Fermi level can lead to substantial changes in conductivity, driven by the complex topology of the Fermi surface. In particular, modifications in Fermi surface nesting emerge as a key factor influencing scattering processes and carrier lifetimes. 
These findings offer critical insights into the microscopic mechanisms that govern transport in TaAs and highlight the sensitivity of Weyl semimetals to doping and carrier dynamics.
\end{abstract}

\maketitle

\section{\label{sec:intro}Introduction}

Topological Weyl semimetals (WSMs) are attracting tremendous interest due to their exotic electronic, optical, and magnetic properties~\cite{Hasan2021, Lv2021, Witczak2012, Yan2017}, as well as their potential applications in quantum technologies~\cite{Hasan2015, Jia2016, Burkov2016, Hu2019, Lu2015}. 
These materials feature nontrivial topological band structures, hosting pairs of Weyl points with opposite chirality where the conduction and valence bands cross linearly, forming a three-dimensional (3D) analog of graphene~\cite{Wan2011}.
Additionally, akin to the surface states in topological insulators, WSMs exhibit Fermi arc surface states, which connect the surface projections of the Weyl points. 
These surface states are a direct consequence of the fact that the Weyl nodes act as topological monopoles, serving as either ``sources'' or ``sinks'' of Berry curvature~\cite{Hasan2017}. 
Taken together, the interplay between bulk Weyl fermions and surface Fermi arcs gives rise to a range of novel physical phenomena, including the chiral anomaly~\cite{Parameswaran2014, Huang2015, Zhang2016}, negative magnetoresistance~\cite{Huang2015, Arnold2016, Li2017, Du2016}, the anomalous quantum Hall effect~\cite{Sodemann2015, Yang2011}, and topological superconductivity~\cite{Meng2012, Hosur2014, Cho2012, Bednik2015, Li2018}. 

Despite extensive experimental efforts, the first Weyl semimetal was only realized in 2015~\cite{Xu2015, Lv2015, Xu2015b, Xu2016, Xu2015c, Yang2015}, following the prediction of a topological Weyl semimetal state in the TaAs materials class~\cite{Weng2015, Huang2015weyl}. 
Angle-resolved photoemission spectroscopy (ARPES) and microwave transmission measurements~\cite{Lv2015prx, Lv2015, Xu2015, Lu2015, Yang2015} confirmed the existence of 12 pairs of Weyl nodes near the Fermi level in the bulk, along with Fermi arc states on the surface. 
Subsequent experimental studies revealed exceptional transport properties in these materials, including ultra-high carrier mobility~\cite{shekhar2015, sankar2017},  extremely large linear magnetoresistance~\cite{Zhang2015, Huang2015weyl}, and chiral-anomaly-induced negative longitudinal magnetoresistance~\cite{Arnold2016, Huang2015, Naumann2020, Zhang2016}. 
Notably, TaAs, the prototypical compound of this family, has been reported to exhibit ultra-high electron mobility of approximately $1.8\times 10^5$~cm$^2$V$^{-1}$s$^{-1}$ at 10~K~\cite{Huang2015} and $4.8 \times 10^5$~cm$^2$V$^{-1}$s$^{-1}$ at 2~K~\cite{Zhang2017}. 
The electron mobility changes rapidly with increasing temperature, dropping by two orders of magnitude above 80~K~\cite{Huang2015}. 
Electron-phonon (e-ph) interactions play a crucial role in determining electronic conductivity in TaAs, but theoretical studies have so far focused on their effects on the phonon linewidth~\cite{Coulter2019}, linear optical conductivity~\cite{Garcia2020}, and thermal transport~\cite{Han2023}. 
The study of carrier transport is further complicated by the material's band topology, where linear band dispersion at the Weyl pockets requires extremely dense Brillouin zone sampling to accurately capture contributions from these localized hot spots, making the computations very challenging.

In this work, we present a first-principles investigation of the temperature-dependent conductivity of TaAs, focusing on the effect of electron-phonon interactions. 
Our computational results, based the iterative solution of the Boltzmann transport equation (BTE), show excellent agreement with experimental measurements. 
By analyzing electron-phonon scattering rates, we identify phonon wave vectors associated with pronounced Fermi surface nesting as the dominant contributors to scattering, whereas the electron-phonon matrix elements remain of the same order of magnitude across the entire Brillouin zone. 
Furthermore, we demonstrate that while the conductivity decreases with electron doping, it is enhanced by hole doping. This behavior is attributed to the evolution of the Fermi surface rather than changes in the electron-phonon coupling strength. 
Finally, calculations using the constant relaxation time approximation (CRTA) exhibit large discrepancies with experiments, highlighting the importance of accounting for the temperature and doping dependence of scattering rates.  
These findings provide valuable insights into the microscopic mechanisms governing transport in TaAs.

\section{Methods}\label{sec:methods}

We use the \textsc{Quantum ESPRESSO} (QE) package~\cite{Giannozzi2017} with optimized norm-conserving Vanderbilt pseudopotentials (ONCVPSP)~\cite{Hamann2013} from the Pseudo Dojo library~\cite{Vansetten2018} generated with the fully relativistic Perdew-Burke-Ernzerhof parametrization~\cite{Perdew1996}. 
We use a plane-wave cutoff of 80~Ry, a Methfessel-Paxton smearing~\cite{Methfessel1989} value of 0.01~Ry, and a $\Gamma$-centered $12^3$ $\bk$-grid. 
The lattice parameters and atomic positions are relaxed until the total energy is converged within $10^{-6}$~Ry and the maximum force on each atom is less than $10^{-4}$~Ry/\AA. 
Spin-orbit coupling (SOC) is included in the electronic structure calculations.
The dynamical matrices and the linear variation of the self-consistent potential due to atomic displacements are calculated within density-functional perturbation theory~\cite{Baroni2001} on a $4^3$ $\mathbf{q}$-grid. 

To investigate electron-phonon interactions and transport properties, we use the EPW code~\cite{Giustino2007,Ponce2016,Lee2023}. 
The electronic wavefunctions required for the Wannier interpolation~\cite{Marzari2012, Pizzi2020} are obtained on a uniform $\Gamma$-centered $8^3$ $\bk$-grid. 
We use 32 atom-centered orbitals to describe the electronic structure of TaAs, using five $d$ and three $p$ orbitals on each Ta and As atom as initial guesses for the Wannier functions, respectively.
The Weyl node positions are determined using the Nelder-Mead algorithm as implemented in WannierTools~\cite{Wu2018}. 

To converge the Fermi level within 0.1~meV, Brillouin zone $\bk$-grids with at least 80$^3$ points are required, as shown in Fig.~S1(b)~\cite{SI}.
This level of accuracy is essential in TaAs because the relative energy positions of the Weyl nodes with respect to the Fermi level are in the order of just a few meV.
Accordingly, we obtain a converged Fermi level, $\ef^0=17.0633$~eV, using the Wannier-interpolated electronic structure within the \textsc{EPW} code~\cite{Ponce2016,Lee2023}, with $\bk$-grids ranging from $80^3$ to $120^3$ and a Fermi-Dirac smearing of 26~meV (corresponding to 300~K). 
The resulting $\ef^0$ is used both to locate the Weyl points and in the transport calculations for the undoped system.

The semimetallic nature of TaAs, combined with its complex Fermi surface featuring multiple electron- and hole-like pockets arising from both Weyl nodes crossings and trivial states, makes transport calculations particularly sensitive to the Fermi level position and the sampling density of the $\bk$- and $\bq$-point grids. 
The Boltzmann transport equations are solved on fine uniform $140^3$ $\bk$- and $70^3$ $\bq$-point grids, with an energy window of $\pm 0.2$~eV around the Fermi level. 
These grids yield well-converged results, as demonstrated by our convergence tests in Fig.~S3(a)~\cite{SI}. 
%
%
All Dirac $\delta$ functions appearing in the expressions for the iterative BTE, scattering rate, phonon linewidth, and nesting function are approximated by Gaussians with a broadening of 2~meV. 

To evaluate the variation in electron-phonon coupling among different phonon modes, we define the modulus of the mode-resolved e-ph coupling strength as
\begin{equation}\label{eq:gqnu}
    |g_{\bq \nu}| = \left[\sum_{mn\bk}|g_{m n \nu}(\bk, \bq)|^2/(N_{\bk} N_{\rm b}^2)\right]^{1/2},
\end{equation}
where $g_{m n \nu}(\bk, \bq)$ is the e-ph coupling matrix element for the scattering between the electronic states $\ket{n \bk}$ and $\ket{m \bk+\bq}$ through a phonon of frequency $\omega_{\bq \nu}$. Here, $N_{\bk}$ is the number of $\bk$-points and $N_{\rm b} = 8$ is the number of bands near the Fermi level included in the summation.

To elucidate the role of e-ph interactions, we also compute the phonon linewidth 
\begin{multline} \label{eq:linewidth}
\g_{\bq \nu} = \pi \sum_{mn}\int \frac{d\bk}{\Omega^{\rm BZ}}|g_{m n \nu}(\bk, \bq))|^2    \\
 \times (f^0_{n\bk}- f^0_{m \bk+\bq}) \d(\ve_{m \bk+\bq} - \ve_{n\bk}-\hbar \o_{\bq\nu}),
\end{multline}
and the Fermi surface nesting function
\begin{multline}\label{eq:nesting}
\zeta_{\bq} = \sum_{m n \nu} \int \frac{d\bk}{\Omega^{\rm BZ}} \frac{1}{\hbar\omega_{\bq\nu}} (f^0_{n\bk}- f^0_{m \bk+\bq}) \\
\times\d(\ve_{m \bk+\bq} - \ve_{n\bk}-\hbar\o_{\bq\nu}).
\end{multline}
At low temperatures, the Fermi factors restrict the $\bk$ integration to a narrow region about the Fermi surface of width $\o_{\bq\nu}$, reducing the phonon linewidth and the nesting function to the following expressions~\cite{Allen1972}:
\begin{align} \label{eq:linewidth-delta}
\g_{\bq \nu} =& \, \pi \hbar\o_{\bq\nu} \sum_{mn}\int \frac{d\bk}{\Omega^{\rm BZ}}|g_{m n \nu}(\bk, \bq))|^2   \nonumber \\
 & \times \d(\ve_{n \bk} - \ve_{\rm F})\d(\ve_{m \bk+\bq} - \ve_{\rm F}), \\
\zeta_{\bq} =& \sum_{m n \nu} \int \frac{d\bk}{\Omega^{\rm BZ}}
\d(\ve_{n \bk} - \ve_{\rm F}) 
\d(\ve_{m \bk+\bq} - \ve_{\rm F}).\label{eq:nesting-delta}
\end{align}

The electrical conductivity is calculated by solving the linearized Boltzmann transport equation~\cite{Ponce2020}:
\begin{equation}\label{eq:sigma}
\sigma_{\alpha\beta} = -\frac{e}{(2\pi)^3} \sum_{n} \int d\bk \,
v_{n\bk, \alpha} \partial_{E_{\beta}} f_{n\bk},
\end{equation}
where $e$ is the electronic charge, and $v_{n\bk, \alpha}$ is the band velocity of an electron with momentum $\bk$ at band $n$ along the Cartesian direction $\alpha$. 
\begin{figure*}[t]
    \centering
    \includegraphics[width=\textwidth]{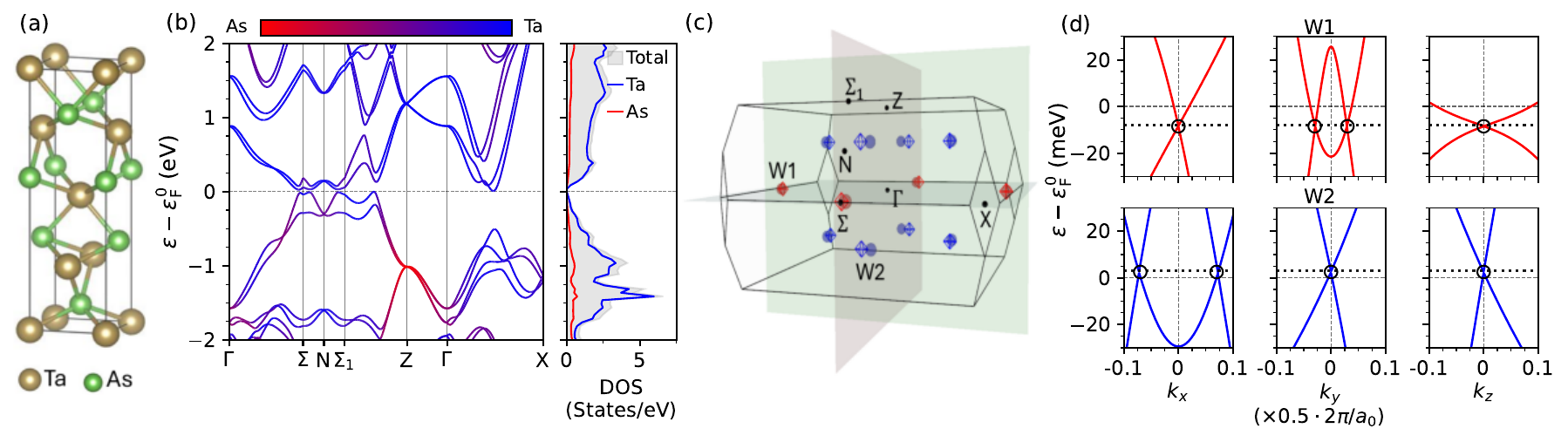}
    \caption{(a) Crystal structure of TaAs, with Ta and As atoms shown as brown and green spheres, respectively. 
    (b) Electronic band structure with orbital-resolved (fat band) character, along with total and partial density of states (DOS). (c) Brillouin zone with the $k_x=0$ and $k_y=0$ mirror planes showing the 12 pairs of Weyl nodes. 
    The two distinct types of Weyl nodes, W1 and W2, are indicated by red and blue colors, respectively. Solid spheres and open diamonds represent the opposite chirality of the Weyl points. 
    (d) Zoomed-in view of the band structure highlighting the Weyl nodes W1 located approximately 8.5~meV below the Fermi level and the Weyl nodes W2 located approximately 2.5~meV above it. The $a_0$ in the $x$-axis label denotes the unit cell edge length, given by $\sqrt{2a^2+c^2}/2$.} 
    \label{fig:band}
\end{figure*}
The function $f_{n\bk}$ represents the electronic occupation factor, and its derivative with respect to the electric field, $\partial_{E_{\beta}} f_{n\bk}$, is obtained by iteratively solving the following equation:
\begin{align} \label{eq:bte}
\tau_{n\bk}^{-1} &\partial_{E_{\beta}} f_{n\bk} = 
e \frac{\partial f_{n\bk}^0}{\partial\ve_{n\bk}}
v_{n\bk, \beta} +
\frac{2 \pi}{\hbar} \sum_{m\nu}
\int \frac{\dd\bq}{\Omega_{\rm BZ}}
|g_{m n \nu}(\bk, \bq)|^2   \nonumber \\
&\times \left [ (n_{\bq \nu} + 1 - f^0_{n \bk})
\d(\ve_{n\bk} - \ve_{m \bk+\bq} + \hbar \o_{\bq\nu}) \right. \nonumber \\
&\left. \, + \, 
(n_{\bq \nu}+ f^0_{n \bk})
\d( \ve_{n\bk} - \ve_{m \bk+\bq} -\hbar \o_{\bq\nu}) \right]
\partial_{E_{\beta}} f_{m\bk+\bq}.
\end{align}
Here, $\ve_{n\bk}$ is the energy eigenvalue of the electronic state $\ket{n \bk}$, $\Omega^{\rm BZ}$ is the volume of the Brillouin zone, $f^0_{n\bk}$ is the Fermi-Dirac distribution function, and $n_{\bq \nu}$ is the bosonic occupation factor. 
The scattering rate $\tau_{n\bk}^{-1}$ is defined as
\begin{multline} \label{eq:tau}
\tau_{n\bk}^{-1} =  \frac{2 \pi}{\hbar} \sum_{m\nu}
\int \frac{\dd\bq}{\Omega^{\rm BZ}}
|g_{m n \nu}(\bk, \bq))|^2    \\
 \, \times \left[ (n_{\bq \nu} + 1 - f^0_{m \bk+\bq})
\d(\ve_{m \bk+\bq} - \ve_{n\bk} + \hbar \o_{\bq\nu}) \right.  \\
\left.  + (n_{\bq \nu}+ f^0_{m \bk+\bq}) \d(\ve_{m \bk+\bq} - \ve_{n\bk}-\hbar \o_{\bq\nu}) \right].
\end{multline}
Omitting the second term on the right-hand side of Eq.~\eqref{eq:bte} reduces the expression for $\sigma_{\alpha\beta}$ to the self-energy relaxation time approximation (SERTA) form of the conductivity~\cite{Ponce2021}:
\begin{align}\label{eq:sigma_SERTA}
\sigma_{\alpha\beta}^{\rm SERTA} = 
-\frac{e^2}{(2\pi)^3} \sum_{n} \int d\bk \,
\frac{\partial f_{n\bk}^0}{\partial\ve_{n\bk}}v_{n\bk, \alpha} v_{n\bk, \beta}\tau_{n\bk}.
\end{align}
Furthermore, if the relaxation time in Eq.~\eqref{eq:sigma_SERTA} is approximated as a constant, $\tau_{n\bk} = \tau$, the expression takes the form used in the CRTA.

To analyze the scattering rates in detail, we also define the energy-averaged scattering rate
\begin{align}\label{eq:tau1avg}
\tau^{-1}(\ve) = \frac{\sum_{n\bk}\tau^{-1}_{n\bk}\d(\ve_{n \bk} - \ve) }
{\sum_{n\bk}\d(\ve_{n \bk} - \ve)}.
\end{align}
and the mode-resolved scattering rate~\cite{Ponce2019}
\begin{multline} \label{eq:tau_domega}
    \frac{\partial \tau_{n\bk}^{-1}}{\partial \omega} =  \,
    \frac{2 \pi}{\hbar} \sum_{m\nu}
    \int \frac{\dd\bq}{\Omega^{\rm BZ}}
    |g_{m n \nu}(\bk, \bq))|^2  \, \delta(\omega_{\bq \nu} - \omega) \\
     \, \times \left[ (n_{\bq \nu} + 1 - f^0_{m \bk+\bq})
    \d(\ve_{m \bk+\bq} - \ve_{n\bk} + \hbar \o_{\bq\nu}) \right. \\
    \left. \, + \,
    (n_{\bq \nu}+ f^0_{m \bk+\bq})
    \d(\ve_{m \bk+\bq} - \ve_{n\bk}-\hbar \o_{\bq\nu}) \right].
\end{multline}
Note that the difference between Eq.~\eqref{eq:tau_domega} and Eq.~\eqref{eq:tau} is the $\delta(\omega_{\bq \nu} - \omega)$ term in the former, such that
\begin{equation}\label{eq:intmode-taunk}
    \tau_{n\bk}^{-1} = \int \! \dd\omega \, \frac{\partial \tau_{n\bk}^{-1}}{\partial \omega} .
\end{equation}
We can also define the mode-resolved energy-averaged scattering rate as
\begin{equation}\label{eq:mode-tau}
    \frac{\partial \tau^{-1} (\ve)}{\partial \omega}=\frac{\sum_{n\bk}\frac{\partial \tau_{n\bk}^{-1}}{\partial \omega}\delta(\ve_{n\bk} - \ve)}
    {\sum_{nk}\delta(\ve_{n\bk} - \ve)}.
\end{equation}
Similar to Eq.~\eqref{eq:intmode-taunk}, we have
\begin{equation}
    \tau^{-1}(\ve) = \int \! \dd\omega \, \frac{\partial \tau^{-1}(\ve)}{\partial \omega} .
\end{equation}

\section{Results and discussion}\label{sec:results}

TaAs crystallizes in the body-centered tetragonal lattice (space group $I4_1md$, no.~109) with broken spatial inversion symmetry~\cite{Weng2015, Huang2015weyl}, a key condition for the realization of Weyl fermions. 
The structure comprises interpenetrating Ta and As sublattices, shifted relative to each other by ($a/2,\, a/2,\, c/12$), as illustrated in Fig.~\ref{fig:band}(a). 
The optimized lattice parameters are $a$ = 3.456~\AA{} and $c$ = 11.719~\AA{}, $0.7\%$ larger than the experimentally measured values~\cite{Huang2015, Lv2015prx, Lv2015}, and  in good agreement with previous theoretical studies~\cite{Han2023, Lv2015, Huang2015weyl}.

The electronic and topological structure of TaAs has been extensively studied in previous works~\cite{Huang2015weyl,Weng2015,Lee2015}, so we summarize only the key aspects here. 
\begin{figure*}[!t]
    \centering
    \includegraphics[width=0.99\linewidth]{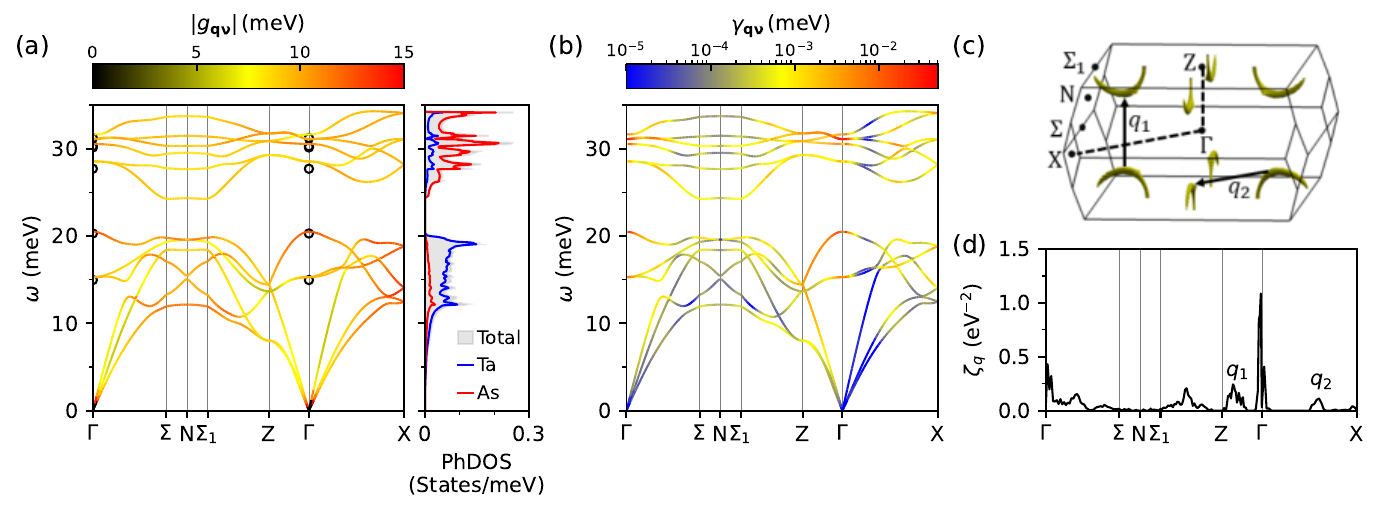}
    \caption{Phonon dispersion of TaAs with (a) mode-resolved electron-phonon coupling strength $|g_{\bq \nu}|$ and (b) phonon linewidths $\gamma_{\bq \nu}$ indicated by a color scale. 
    The total and atom-projected phonon density of states (PhDOS) are shown alongside panel (a). 
    Hollow black circles in the phonon dispersion in panel (a) correspond to Raman data from Ref.~\citenum{Liu2015}. 
    (c) Fermi surface generated using a 10~meV energy window around the Fermi level with the XCrySDen package~\cite{Kokalj1999}. 
    The surface reveals eight banana-shaped hole pockets, each containing a pair of W2 nodes. 
    The black solid arrows indicate possible inter-pocket scatterings with $\bq_1$ and $\bq_2$ wave vectors along the $\Gamma$-$Z$ and $\Gamma$-$X$ directions (dashed lines), respectively. 
    (d) Nesting function shown along the same high-symmetry path as in panels (a) and (b). 
    The two peaks along the $\Gamma$-$Z$ and $\Gamma$-$X$ directions are labeled by wave vectors $\bq_{1}$ and $\bq_{2}$.}
    \label{fig:phband}
\end{figure*}
In the presence of SOC, the band structure becomes fully gapped along all high-symmetry directions. However, the valence band maximum and conduction band minimum lie at the Fermi level ($\ef^0$), resulting in a semimetallic system, as shown in Fig.~\ref{fig:band}(b) (see Fig.~S1(a)~\cite{SI} for a zoomed-in view near $\ef^0$). 
Orbital decomposition reveals that Ta-5$d$ states dominate both sets of bands in this energy region, while As-4$p$ states contribute only minimally. 
Due to the semimetallic character, the density of states (DOS) remains low at $\ef^0$ but increases rapidly with energy into both the valence and conduction bands.

As shown in Fig.~\ref{fig:band}(c), 12 pairs of Weyl nodes emerge in the bulk Brillouin zone due to the broken inversion symmetry~\cite{Weng2015, Huang2015weyl}. 
In the absence of SOC, gapless crossings of the valence and conduction bands along some high-symmetry lines form two closed nodal lines in both the $k_x=0$ and $k_y=0$ mirror planes~\cite{Huang2015weyl}. 
Inclusion of SOC gaps the four nodal lines, each of them evolving into three pairs of gapless points slightly displaced from the mirror planes (one pair in the $k_z=0$ plane, labeled as W1, and two pairs in the $k_z=\pm \pi/c$ planes, labeled as W2, following the convention in Refs.~\citenum{Yan2017, Huang2015}). 
Fig.~\ref{fig:band}(d) presents a zoomed-in view of the band structure, highlighting the two distinct types of Weyl nodes. 
The W1 nodes are located 8.5~meV below the Fermi level, while the W2 nodes remain much closer, 2.5~meV above it, consistent with ARPES measurements~\cite{Zhang2016}.
Moreover, the 11~meV energy offset between the two sets of Weyl nodes is in excellent agreement with prior first-principles calculations (14~meV)~\cite{Xu2015,Lee2015} and ARPES measurements (13~meV)~\cite{Zhang2016}.

\begin{figure}[!t]
    \includegraphics[width=\linewidth]{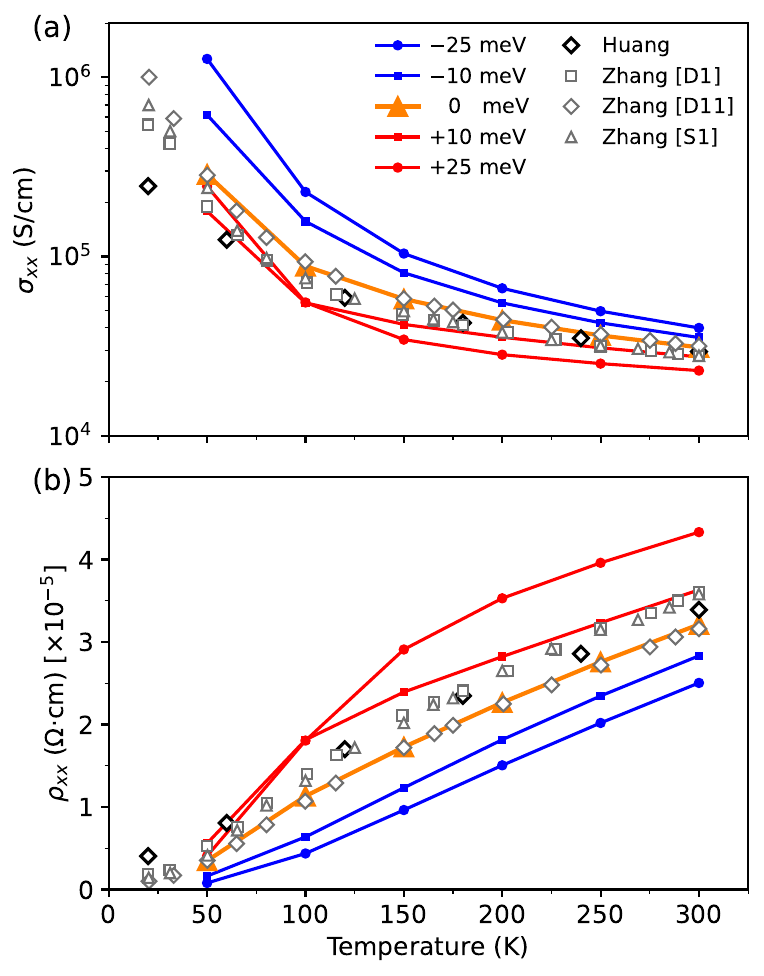}
    \caption{\label{fig:CondT} (a) Electrical conductivity and (b) resistivity of TaAs for the undoped system ($\ef=\ef^0$) and for two levels of electron and hole doping ($\ef = \ef^0 \pm 25$~meV and $\ef = \ef^0 \pm 10$~meV). 
    Experimental data from Refs.~\citenum{Huang2015} and \citenum{Zhang2017} are shown as hollow symbols for comparison. 
    The three highest quality samples, labeled as S1, D11, and D1 in Ref.~\citenum{Zhang2017}, are used. }
\end{figure}

\begin{figure*}[!t]
    \includegraphics[width=0.99\linewidth]{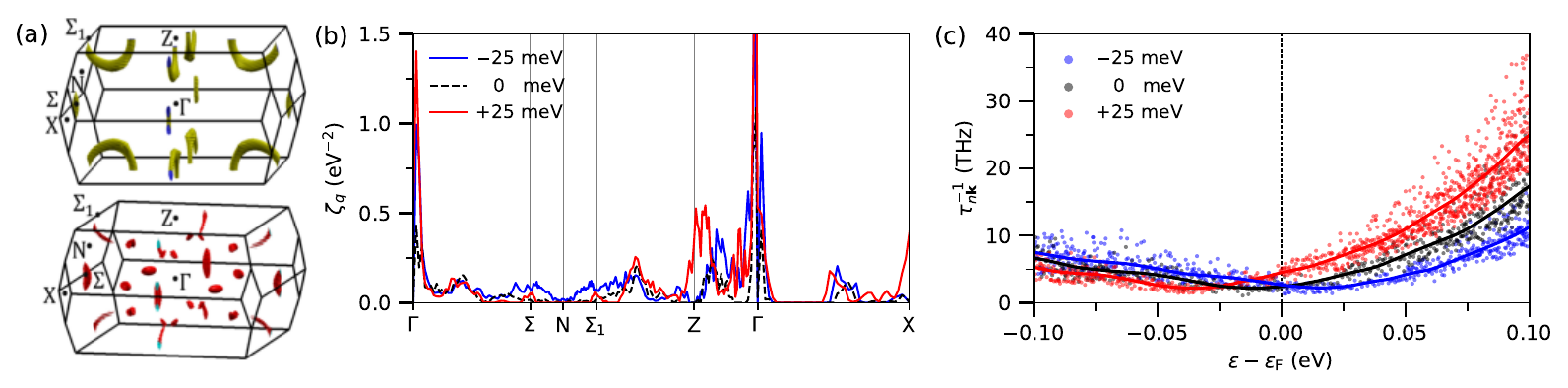}
    \caption{\label{fig:scater} (a) Three dimensional Fermi surface, (b) nesting function, and (c) scattering rates for hole-doped ($\ef = \ef^0-25$~meV) and electron-doped ($\ef = \ef^0 + 25$~meV) TaAs. 
    Panels (b) and (c) show quantities evaluated at 300~K, with results for the undoped system ($\ef=\ef^0$) included for comparison. 
    In panel (c), the energy-averaged scattering rates are shown as solid lines.}
\end{figure*}

Fig.~\ref{fig:phband}(a) illustrates the phonon dispersion of TaAs, color-coded by the e-ph coupling strength, along with the projected phonon density of states (PhDOS). 
The acoustic modes (0--15~meV) are smoothly dispersive and primarily involve Ta contributions, as evidenced by the overlapping purple (Ta) and blue (As) PhDOS projections. As previously noted~\cite{Ouyang2016, Han2023}, the optical phonon spectrum is clearly split into two distinct regions by a gap of approximately 4~meV, indicating weak hybridization between vibrations of the heavier Ta atoms and the lighter As atoms. 
Consistent with their mass difference, the lower-frequency vibrations (15--20~meV) are mainly associated with Ta atoms, while higher-frequency modes (24--35~meV) are dominated by As vibrations. 
Raman measurements confirm all optical phonon modes at the Brillouin zone center, identifying one $A_1$, two $B_1$, and three $E$ modes~\cite{Liu2015}. Their distinct symmetries and energies are consistent with our first-principles results. 

To shed light on the role of the e-ph interactions in TaAs, we evaluate the modulus of the mode-resolved e-ph coupling strength $|g_{\bq \nu}|$, as defined in Eq.~\eqref{eq:gqnu} (see Sec.~\ref{sec:methods}).
Analysis of Fig.~\ref{fig:phband}(a) shows that the coupling strength is relatively uniform across most modes, with an average of around 10~meV. 
The strongest coupling comes from the low-energy optical modes within the 15--20~meV range, reaching values up to 15~meV, as indicated by the deep red coloring at the $\Gamma$ and $Z$ points and along the $\Sigma$-$N$-$\Sigma_1$ and $\Gamma$-$Z$ directions. 
In contrast, the phonon linewidths $\gamma_{\bq \nu}$, defined in Eq.~\eqref{eq:linewidth} and shown in Fig.~\ref{fig:phband}(b), span a much broader range from $10^{-2}$ down to $10^{-5}$~meV. 
The largest values are again associated with the same low-energy optical modes, though now primarily near the $\Gamma$ point and along the $\Gamma$-$Z$ path, consistent with previous findings~\cite{Coulter2019}. 
This wide variation suggests that the linewidths are more strongly influenced by the structure of the Fermi surface than by the strength of the e-ph coupling.

In order to disentangle the two contributions, we further examine the nesting function depicted in Fig.~\ref{fig:phband}(d). 
For consistency with the phonon linewidth analysis, we focus on the results obtained with  Eq.~\eqref{eq:nesting}.
However, similar trends were found using the commonly adopted double-delta approximation given in Eq.~\eqref{eq:nesting-delta} (see Fig.~S2~\cite{SI}). 
The nesting function exhibits its highest peak near the $\Gamma$ point, with pronounced features along the $\Gamma$-$\Sigma$, $Z$-$\Sigma_1$, and $\Gamma$-$Z$ directions, as well as a smaller peak along $\Gamma$-$X$. 
To interpret these features, we show the Fermi surface in Fig.~\ref{fig:phband}(c), which reveals eight banana-shaped hole pockets, each containing a pair of W2 nodes. 
The strong peak in $\zeta_{\bq}$ at $\Gamma$ can be attributed to the intra-pocket scattering, while the other peaks are associated with the inter-pocket scattering. In particular, the black arrows in Fig.~\ref{fig:phband}(c) indicate possible inter-pocket transitions mediated by phonons with wave vectors $\bq_{1}$ and $\bq_{2}$ along the $\Gamma$-$Z$ and $\Gamma$-$X$ directions, respectively. 
These observations highlight the key role of Fermi surface nesting in governing e-ph scattering processes in TaAs. 

We now turn to the transport properties of TaAs and compute the temperature-dependent electrical conductivity using the iterative BTE~\cite{Ponce2020}, as described in Sec.~\ref{sec:methods}. 
Fig.~\ref{fig:CondT}(a) displays the calculated conductivity for the undoped system ($\ef =\ef^0$) and for two levels of electron and hole doping ($\ef = \ef^0 \pm 25$~meV and $\ef = \ef^0 \pm 10$~meV). 
In all cases, the transport behavior is metallic, with the conductivity decreasing as temperature increases. 
We find that the computed conductivity for the undoped system and the lowest level of electron doping closely match experimental data~\cite{Huang2015, Zhang2017}. 
Notably, while measurements for nine samples are reported in Fig. 1c of Ref.~\citenum{Zhang2017}, we only consider the three highest-quality samples with the largest residual resistance ratios, labeled D1, D11, and S1. 
As pointed out in that study, sample quality is highly sensitive to the growth conditions, which in turn strongly influence the position of the Fermi level and, consequently, the electrical conductivity. Our results also reveal a strong sensitivity to the position of $\ef$. 
Specifically, hole doping yields higher conductivity values than electron doping across the entire temperature range. 
For completeness, we also present the resistivity in Fig.~\ref{fig:CondT}(b), which shows that the slope of the temperature dependence of resistivity for the undoped case is close to that observed in experimental samples~\cite{Huang2015, Zhang2017}. 
These results confirm that the transport in high-quality TaAs samples is phonon-limited.

A comparison with results from the SERTA shows that it produces conductivities that differ from the iterative BTE by less than $5\%$ (see Fig.~S3(b)~\cite{SI}).
Given the small variation in e-ph matrix elements across the Brillouin zone, we also computed the conductivity using the CRTA, adopting a fixed relaxation time of $\tau=0.216~$ps at 300~K to reproduce the iterative BTE result in the undoped case.
As shown in Figs.~S3(b) and S4~\cite{SI}, the significant discrepancies between CRTA and iterative BTE at different temperatures or under electron doping indicate that CRTA is inadequate for describing transport in TaAs. 

To understand the observed trends in conductivity, we first analyze how electron and hole doping modify the Fermi surface and nesting function. 
Compared to the undoped system, hole doping preserves the eight original large hole pockets and introduces several additional small hole pockets around the $k_z$ plane, where the W1 Weyl nodes are located (see Fig.~\ref{fig:scater}). 
In contrast, electron doping markedly alters the Fermi surface topology, leading to the disappearance of hole pockets and the emergence of multiple smaller electron pockets. 
As a result, the nesting function undergoes more pronounced changes in the electron-doped case, with additional peaks near the $Z$ and $X$ points and an enhanced peak at $\Gamma$. 
These differences are also reflected in the scattering rates. 
The energy-averaged scattering rates $\tau^{-1}(\ve)$, defined in Eq.~\eqref{eq:tau1avg}, shown as solid lines in Fig.~\ref{fig:scater}, indicate that $\tau^{-1}(\ef)$ remains comparable to the undoped case for hole doping, but increases by almost a factor of two for electron doping. 

Additional insight into this asymmetric behavior can be gained by examining the phonon linewidth differences between the doped and undoped systems, as presented in Fig. S5~\cite{SI}. 
While $\g_{\bq \nu}$ is largely unchanged throughout the Brillouin zone, notable variations appear along the $\Gamma$-$Z$ and $\Gamma$-$X$ directions. 
In the hole-doped case, a strong suppression is observed along $\Gamma$-$Z$, consistent with the small changes in Fermi surface topology and nesting. 
Conversely, electron doping leads to a pronounced enhancement along $\Gamma$-$Z$ and a moderate increase along $\Gamma$-$X$, indicative of stronger e-ph scattering in these momentum regions. 
These enhancements correlate with the more substantial changes in the Fermi surface and nesting function, explaining the higher scattering rates observed under electron doping. 
This analysis is consistent with the fact that CRTA fails to capture the doping dependence of conductivity in the electron-doped case, as shown in Fig.~S4~\cite{SI}.

\begin{figure}[!b]
    \centering
    \includegraphics[width=0.95\linewidth]{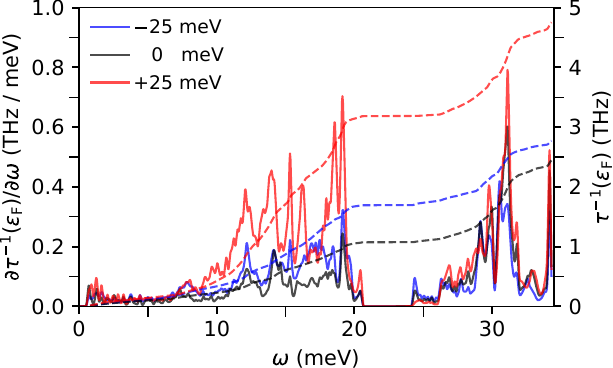}
    \caption{Frequency-resolved scattering rate $\partial \tau^{-1}(\ve)/\partial \omega$ (solid curves, left axis) and cumulative integral $\int_0^{\omega} \rm{d}\omega'\partial \tau^{-1}(\ve)/\partial \omega'$ (dashed curves, right axis) evaluated at the Fermi level for undoped ($\ef = \ef^0$), hole-doped ($\ef = \ef^0 - 25$~meV), and electron-doped ($\ef = \ef^0 + 25$~meV) systems at 300~K.}
    \label{fig:mode-tau}
\end{figure}
We corroborate our analysis on the origin of the enhanced scattering rates in electron-doped systems by further analyzing the contribution of each phonon mode to the scattering rate. 
To this end, we evaluate the electron-energy-averaged, phonon-frequency-resolved scattering rate $\partial \tau^{-1} (\ve)/\partial \omega$, defined in Eq. ~\eqref{eq:mode-tau}.
Fig.~\ref{fig:mode-tau} presents $\partial \tau^{-1}(\ve)/\partial \omega$ and its cumulative integral $\int_0^{\omega} \rm{d}\omega' \partial \tau^{-1}(\ve)/\partial \omega'$, which represents the electron-energy-averaged scattering rate. 
These quantities are evaluated at the Fermi level for three cases, corresponding to $\ef = \ef^0$ and $\ef = \ef^0 \pm 25$~meV. 
All three spectral profiles exhibit similar features, with a broad peak in the mid-frequency region ($10-20$~meV) and a narrower peak in the high-frequency region ($30-35$~meV), both attributed primarily to the optical phonon modes along the $\Gamma$-$Z$ direction (see Fig.~\ref{fig:phband}). 
The pronounced enhancement in the mid-frequency scattering upon electron doping is reflected in the cumulative integral, which approaches 5~THz, nearly twice the value obtained for the undoped and hole-doped systems.
This trend aligns with the increase in the nesting function around the Z point under electron doping (see Fig.~\ref{fig:scater}), and corresponds to the observed conductivity reduction. 
For comparison, the spectral dependence in the hole-doped system remains close to the undoped one, indicating that the increase in conductivity with hole-doping stems primarily from the enhanced DOS at the Fermi level.

\section{\label{sec:summary}Summary}

In summary, using state-of-the-art \textit{ab initio} transport calculations, we have shown that electrical conductivity in high-quality TaAs samples is well described by phonon-limited transport. 
Our results reveal that conductivity is highly sensitive to carrier doping due to the complex topology of the Fermi surface. 
In particular, electron doping leads to a reduction in conductivity relative to the undoped case, which can be attributed to the formation of additional electron pockets that enhance the nesting function and open new scattering channels. 
In contrast, hole doping results in increased conductivity, driven by an enhancement in the density of states at the Fermi level compared to the undoped system. Analysis of the phonon lifetimes and the spectral decomposition of scattering rates further demonstrates that optical phonons along the $\Gamma$-$Z$ direction play a dominant role in charge carrier relaxation. 
These findings provide a fundamental understanding of the electron-phonon coupling mechanisms in TaAs and pave the way for similar investigations in other Weyl semimetals. 



\begin{acknowledgements}
We thank C.-L.~Zhang for providing the experimental raw resistivity data of Ref.~\citenum{Zhang2017} and for helpful discussions. 
S. P. is a Research Associate of the Fonds de la Recherche Scientifique - FNRS.
This work was primarily supported by the Computational Materials Sciences Program funded by the U.S. Department of Energy, Office of Science, Basic Energy Sciences, under Award No. DE-SC0020129.
This work was also supported by the Fonds de la Recherche Scientifique - FNRS under Grants number T.0183.23 (PDR) and  T.W011.23 (PDR-WEAVE). 
This publication was supported by the Walloon Region in the strategic axe FRFS-WEL-T.
The authors acknowledge the computational resources provided by the Frontera and Stampede3 supercomputers at the Texas Advanced Computing Center (TACC) at The University of Texas at Austin (http://www.tacc.utexas.edu), supported through the Leadership Resource Allocation (LRAC) award DMR22004 and the ACCESS allocation TG-DMR180071, respectively.

\end{acknowledgements}

\section*{Author Contributions}
Z.L. and S.M. contributed equally. Z.L. and S.M. modified the EPW code and wrote the original draft. Z.L. tested the code and carried out the transport calculations, while S.M. prepared the figures. R.M. conceived and supervised the project, reviewed and edited the manuscript, and secured the funding. All authors participated in the formal analysis and the revision of the paper.

\section*{Data Availability}

The source code and data associated with this work are available in the Materials Cloud Archive~\cite{MaterialCloudArchive}.





%

\end{document}


\title{Supporting Information: \\Phonon-limited carrier transport in the Weyl semimetal TaAs}

\author{Zhe Liu}
\affiliation{Department of Physics, Applied Physics and Astronomy, Binghamton University-SUNY, Binghamton, New York 13902, USA}
\author{Shashi B. Mishra}
\affiliation{Department of Physics, Applied Physics and Astronomy, Binghamton University-SUNY, Binghamton, New York 13902, USA}
\author{Jae-Mo Lihm}
\affiliation{European Theoretical Spectroscopy Facility and Institute of Condensed Matter and Nanosciences (IMCN), Universit\'e catholique de Louvain (UCLouvain), Belgium}
\author{Samuel Ponc\'{e}}
\affiliation{European Theoretical Spectroscopy Facility and Institute of Condensed Matter and Nanosciences (IMCN), Universit\'e catholique de Louvain (UCLouvain), Belgium}
\author{Elena R. Margine}
\affiliation{Department of Physics, Applied Physics and Astronomy, Binghamton University-SUNY, Binghamton, New York 13902, USA}

\date{\today}

\maketitle



\begin{figure}[!hbt]
    \centering
    \includegraphics[width=\textwidth]{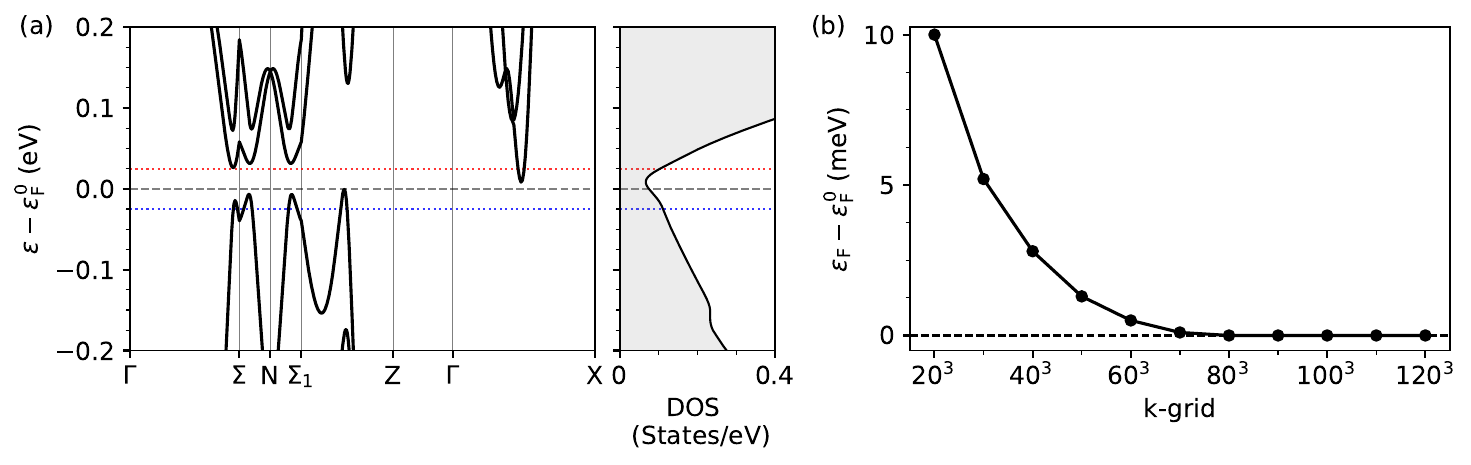}
    \caption{(a) Zoomed-in view of the electronic band structure and DOS of TaAs. The red, black, and blue dashed lines indicate energy levels at 25~meV, 0~meV, and $-$25~meV relative to $\ef^0$, respectively. (b) Convergence of the Fermi level $\ef$ of TaAs with respect to the $\bk$ points grid at 300~K. $\ef^0$ denotes the converged Fermi level obtained for $\bk$ grids in the $80^3-120^3$ range, as indicated by the dashed line.}
    \label{fig:zoom-band}
\end{figure}

\begin{figure}[!hbt]
    \includegraphics[width=0.45\textwidth]{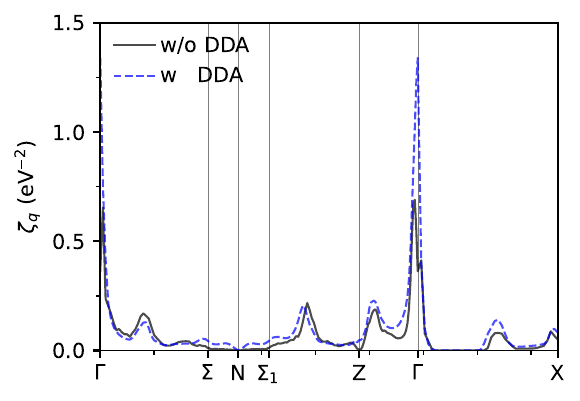}
    \caption{\label{fig:doubledelta} Comparison of the nesting function computed without (w/o DDA) and with (w DDA) the double-delta approximation at 300~K for the undoped system.}
\end{figure}

\begin{figure}[!hbt]
    \includegraphics[width=\textwidth]{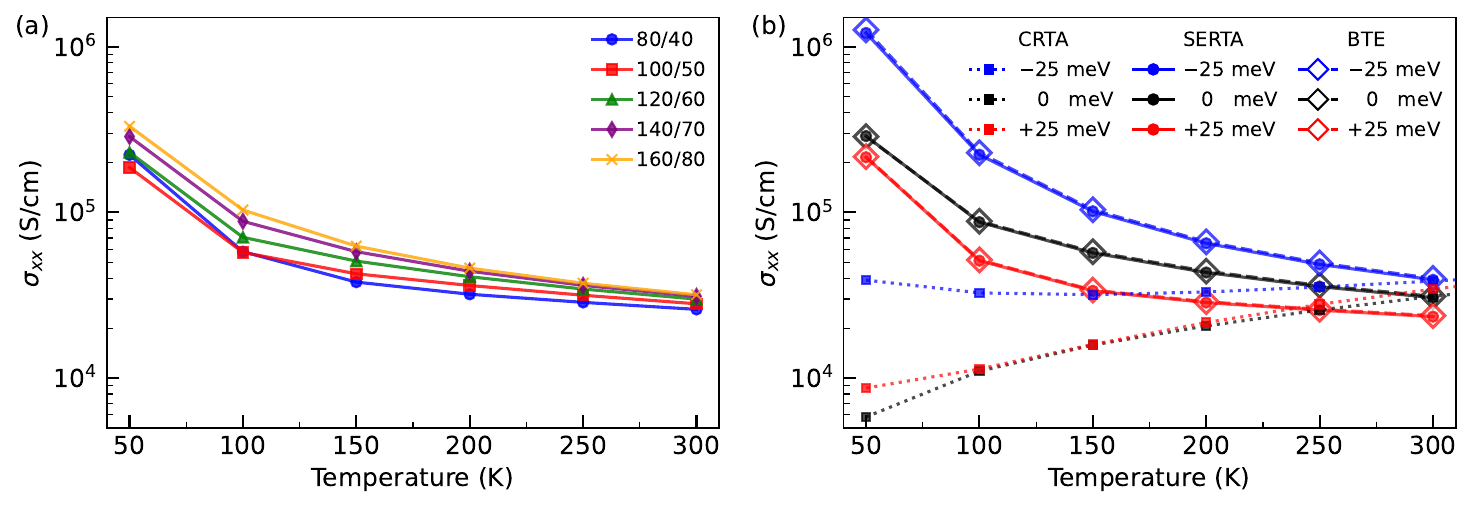}
    \caption{\label{fig:Conv-grid} (a) Convergence of the electrical conductivity computed using the iterative BTE with respect to \textbf{k}/\textbf{q}-point grids at 300~K for the undoped system. 
    (b) Comparison of conductivity results obtained using the iterative BTE, SERTA, and CRTA for the undoped system ($\ef=\ef^0$) and one level of electron and hole doping ($\ef = \ef^0 \pm 25$~meV) at different temperatures, computed on $140^3$ $\bk$- and $70^3$ $\bq$-point grids.} 
\end{figure}
\begin{figure}[!hbt]
    \includegraphics[width=0.5\textwidth]{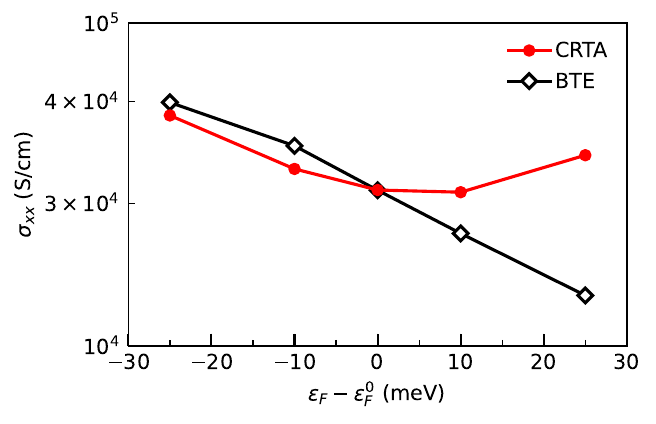}
    \caption{\label{fig:CRTA} Comparison of conductivity results computed using the iterative BTE and CRTA at 300~K as a function of the Fermi level. The undoped case corresponds to $\ef = \ef^0$, while electron and hole doping are represented by $\ef = \ef^0 \pm 10$~meV and $\ef = \ef^0 \pm 25$~meV, computed on $140^3$ $\bk$- and $70^3$ $\bq$-point grids.} 
\end{figure}
\begin{figure}[!hbt]
    \includegraphics[width=0.75\textwidth]{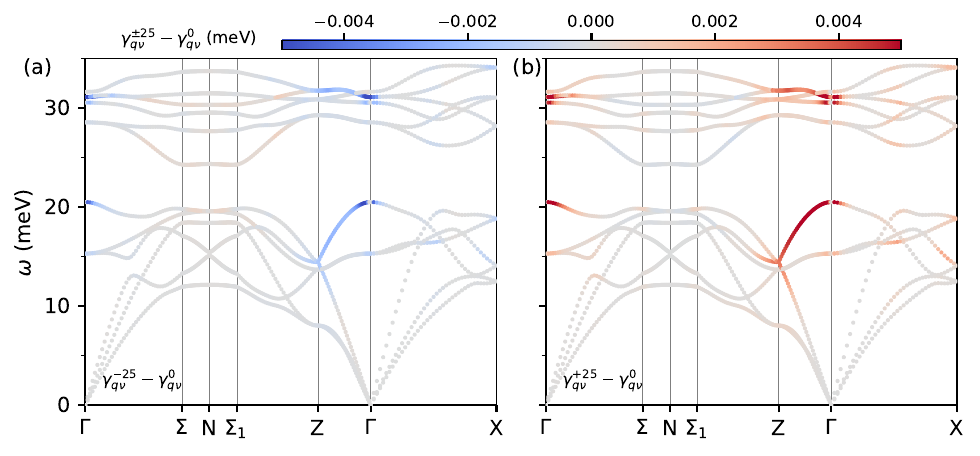}
    \caption{\label{fig:gamma-doped} Difference in phonon linewidths $\gamma_{\textbf{q} \nu}$ between the doped and undoped ($\ve_{\rm F} = \ve_{\rm F} ^0$) systems: (a) hole-doped ($\ve_{\rm F} = \ve_{\rm F} ^0 - 25$~meV) and (b) electron-doped ($\ve_{\rm F} = \ve_{\rm F} ^0 + 25$~meV) at 300~K. A significant enhancement in  $\gamma_{\textbf{q} \nu}$ is observed along the $\Gamma-Z$ direction and, to a lesser extent, along $\Gamma-X$ for electron-doping, whereas hole doping results suppressed linewidths along these paths. }
\end{figure}
%
